\documentclass[twocolumn,prd,showpacs]{revtex4}
\usepackage{graphicx}
\usepackage{amssymb}
\begin{document}

\title{\bf Orbits of particles and black hole thermodynamics in a spacetime with torsion }
\author{S. Gharibi Nodijeh}
\email{s.gharibi96@stu.gu.ac.ir}

\author{S. Akhshabi}
\email{s.akhshabi@gu.ac.ir}

\author{F. Khajenabi}
\email{f.khajenabi@gu.ac.ir}

\affiliation{Department of Physics, Faculty of Sciences, Golestan
University\\ Gorgan, IRAN} \vspace{1cm} 

\begin{abstract}

We derive the static spherically symmetric vacuum solution for a
spacetime with non-vanishing torsion by solving the field equations
analytically. The effects of torsion appear as a single parameter in
the line element. For the positive values of this parameter, the
resulting line element is found to be of the Reissner-Nordstrom
type. This parameter is related to the spin of matter and acts as a
torsion 'charge' much like the electric charge in conventional
Reissner-Nordstrom geometry. We also analyze the existence and
stability of the orbits for both massless and massive particles in
this setup and compare the results to the corresponding case in
general relativity. We also derive the first law of black hole
thermodynamics for a black hole with torsion and define the black
hole temperature and entropy in terms of its mass and torsion
charge.
\end{abstract}

\maketitle
\section{Introduction}

The first hint that black holes may behave as thermodynamical
systems came from Bekenstein \cite{Bekenstein1,Bekenstein2,
Bekenstein3,Bekenstein4}. He suggested that a black hole may have an
entropy proportional to its surface area, however he was not able to
determine the exact relation. A little later, Bardeen, Carter, and
Hawking discovered that there is analogy between the laws of black
hole mechanics and four laws of classical thermodynamics, but with
certain limitations, as in order for the laws of black hole
mechanics to be true thermodynamical equations, black holes ought to
have non-zero temperatures. This was thought to be impossible at the
time because black holes could only absorb and not emit particles
and radiation \cite{bardin}. This conundrum was solved when by
applying quantum field theory to black holes, Hawking realized that
they indeed emit thermal radiation and as a result have a non-zero
temperature related to their surface gravity $\kappa$ by the
relation $T=\frac \kappa {2\pi}$, \cite{Hawking1,Hawking2}. This
discovery led to the famous Hawking - Bekenstein formula for the
black hole Entropy $\Sigma=\frac A 4$, where $A$ is the surface area
of the event horizon (In the above and subsequent relations we use
Planck units $\hbar=c=G=4\pi \epsilon_{0}=1$). This relation has
profound theoretical implications, as it relates quantum mechanical
effects through the entropy to gravity and can lead us to the
ultimate quantum theory of gravity.

On the other hand, torsion appears naturally in many proposed
theories of quantum gravity \cite{Schrek, Hehl1, Hehl2, Shapiro}.
More specifically the low energy effective Lagrangian of string
theory has shown to be equivalent  to a Brans-Dicke generalization
of a metric theory of gravity with torsion \cite{Hammond1,
Hammond2}. It is Also well known that the corresponding field
strength of the Kalb-Ramond field in string theory can act like a
torsion field in the background geometry \cite{Kalb}. The
Kalb-Ramond field also appears in noncommutative field theory
\cite{Seiberg} where the presence of torsion is also established
\cite{Chamseddine}.

By the above arguments, studying the black hole thermodynamics in
the presence of torsion seems to be a matter of interest when one
wants to look for quantum effects in gravity. There are several
important previous works in this subject. In \cite{Blag1} the role
of torsion in three-dimensional quantum gravity is investigated by
studying the partition function of the Euclidean theory in
Riemann-Cartan spacetime. In \cite{Day}, the formalism of spacetime
thermodynamics was extended to Einstein-Cartan theory of gravity. In
\cite{Chakraborty} it has been shown that the presence of spacetime
torsion does not affect the entropy-area relation. In the present
paper we study orbits of particles around a black hole and black
hole thermodynamics in gauge theories of gravity were torsion is
present. These theories are of a great importance from a field
theoretical point of view as they involve localization of spacetime
symmetries, much like the localization scheme of internal symmetries
present in the standard model of particle physics. Gauge theories of
gravity typically involve spin of particles in the gravitational
interactions and as a result may provide a convenient way to study
some quantum effect on gravitational phenomena, specially at high
energies.

The most general gauge theories of gravity are equipped with a
metric and a general linear connection and are usually called
'metric-affine' theories of gravity and are described by a
$(L_{4},g)$ space. If we introduce the metricity condition in these
theories, we will get the Riemann-Cartan space $U_{4}$ and the
resulting theory is called the Poincar{\'e} gauge theory of gravity
(PGT) which applies the localization scheme to the global
Poincar{\'e} group of transformations \cite{blag,Hayashi}. The
dynamical variables in this theory are tetrad and spin connection
fields and the associated field strengths are curvature and torsion
tensors, which are coupled to energy-momentum and spin-density
tensors respectively. There are several important special cases of
PGT, namely General relativity (vanishing torsion), teleparallel
theory (vanishing curvature) and Einstein-Cartan theory where the
Lagrangian is set to be equal to the Einstein-Hilbert Lagrangian of
general relativity. Einstein-Cartan theory offers the simplest
generalization of general relativity and has been studied
extensively in literature \cite{Kerlick}. In this case the torsion
is completely determined by the spin density tensor and can not
propagate \cite{Rauch}. However, propagating torsion modes can be
present in Poincar{\'e} gauge theory of gravity with general
quadratic Lagrangian \cite{Blag2}.

In this paper we study black hole thermodynamics and particle orbits
in Poincar{\'e} gauge theory of gravity. The organization of the
paper is as follows: In section II, we briefly review the gauge
theories of gravity with torsion and present the main equations for
Poincar{\'e} gauge theory of gravity. In section III, static
spherically symmetric solutions to the Poincar{\'e} field equations
was derived. In section IV, the effective potential and orbits of
particles in a static spherically symmetric spacetime with torsion
is studied for  photons and massive particles. Section V devotes to
black hole thermodynamics in the presence of torsion. Finally a
brief review and discussion of the main results is given in the
conclusion.

\section{Gauge theories of gravity with torsion}

In PGT the gravitational field is described by both curvature and
torsion tensors. These in turn can be expressed in terms of tetrad
and spin connection as

 $$\label{ToRiem}
   \,\,{R_{\mu\nu\,
   i}}^j=2\Big({\partial_{[\mu}T_{\nu]i}}^j+{\Gamma_{[\mu|k}}^j
    {\Gamma_{|\nu]i}}^k\Big)$$
\begin{equation}
{T_{\mu\nu}}^i=2\Big({\partial_{[\mu}e_{\nu]}}^i+{\Gamma_{[\mu|j}}^i
{e_{|\nu]}}^j\Big)\,\,,\,\, T_\mu={T_{\mu\nu}}^\nu
\end{equation}

where $e_{~\mu}^{i}$ is the tetrad field and

\begin{equation}
g_{\mu\nu}=\eta_{ij}e_{~\mu}^{i}e_{~\nu}^{j},
\end{equation}

is the spacetime metric. The spin connection is related to the usual
holonomic connection by the relation

\begin{equation}
\Gamma^{~~\nu}_{i
\mu}=e_{~\mu}^{j}e^{~\nu}_{k}\Gamma^{~~~k}_{ij}+e_{~\mu}^{j}\partial_{i}e^{~\nu}_{j}
\end{equation}

 Here the Greek indices refer to holonomic coordinate
bases and the Latin indices refer to the Local Lorentz frame. The
most general Lagrangian of the theory is a quadratic function built
by irreducible decompositions of curvature and torsion. Here we
assume a Lagrangian in the form

\begin{equation}\label{lag}
L_{g}=-\frac {a_0} 2R+\frac b {24}R^2+\frac {a_1}
8\Big(T_{\nu\sigma\mu}T^{\nu\sigma\mu}+2T_{\nu\sigma\mu}T^{\mu\sigma\nu}-4T_\mu
T^\mu\Big)
\end{equation}

Where $a_0$, $a_{1}$ and  $b$ are coupling constants and $R$ is the
Ricci scalar. The field equations then is given by the variation of
the Lagrangian with respect to the tetrad and spin connection fields
and have the general form \cite{Shie}

\begin{eqnarray}
  \nabla_{\nu}H_{i}^{\mu\nu}-E_{i}^{~\mu}&=&{\cal T}_{i}^{~\mu},\\
  \nabla_{\nu}H_{ij}^{~~\mu\nu}-E_{ij}
  ^{~~\mu}&=&S_{ij}^{~~\mu},
 \end{eqnarray}

 where

 \begin{eqnarray}
  H_{i}^{~~\mu\nu}&:=&{\partial e L_{\rm G}\over \partial\partial_{\nu} e_{\mu}^i}
  =2{\partial e L_{\rm G}\over \partial T_{\nu\mu}{}^i},\\
  H_{ij}{}^{\mu\nu}&:=&{\partial e L_{\rm G}\over
   \partial\partial_{\nu}\Gamma_{\mu}^{~ij}}
  =2{\partial e L_g\over \partial R_{\nu\mu}{}^{ij}},
 \end{eqnarray}

 and

 \begin{eqnarray}
  E_{i} {}^{\mu}&:=&e^{\mu}{}_{i} e L_{\rm G}-T_{i \nu}{}^{j} H_{j} {}^{\nu\mu}
  -R_{i\nu}{}^{jk}H_{jk}{}^{\nu\mu},\\
  E_{ij}{}^{\mu}&:=&H_{[ij]}{}^{\mu}.
 \end{eqnarray}

The source terms here are energy-momentum and spin density tensors
respectively and are defined by

\begin{eqnarray}
 {\cal T}_{i}{}^{\mu}&:=&\frac{\partial eL_{\rm M}}{\partial e_{\mu}{}^i},\\
S_{ij}{}^{\mu}&:=&
    \frac{\partial eL_{\rm M}}{\partial \Gamma_{\mu}{}^{ij}},
\end{eqnarray}

where $L_{M}$ is the matter Lagrangian and $e$ is the determinant of
the tetrad.

\section{Solving Field Equations}

First we derive the static spherically symmetric vacuum solutions to
the Poincar{\'e} field equations. This solution would describe the
spacetime outside a static spherically symmetric black hole with
torsion. The corresponding solution in general relativity is the
Schwarzschild metric. For a Lagrangian in the form of (\ref{lag}),
the explicit form of the field equations are \cite{Shie}

\begin{equation}\label{field eq1}
    \bar\nabla_\mu R+\frac2 3\Big(R+\frac{6\mu}b\Big)T_\mu=0
\end{equation}

$$a_0\Big(\bar R_{\mu\nu}-\frac 1 2g_{\mu\nu}\bar R\Big)-\frac b 6R\Big(R_{(\mu\nu)}-\frac 1
4g_{\mu\nu}R\Big)$$
\begin{equation}\label{field eq2}
    -\frac {2\mu}
3\Big(\bar\nabla_{(\mu}T_{\nu)}-g_{\mu\nu}\bar\nabla_\rho
T^\rho\Big)-\frac \mu 9\Big(2T_\mu T_\nu+g_{\mu\nu}T_\rho
T^\rho\Big)=-\tau_{\mu\nu}
\end{equation}

Where $\mu:=a_{1}-a_{0}$, $\bar R_{\mu\nu}$ and $\bar R$ are the
Riemannian Ricci tensor and scalar respectively and $\bar\nabla$ is
the covariant derivative with respect to the Levi-Civita connection
of GR. For black holes, the most general static line element with
desired symmetries is

\begin{equation}\label{metric}
    {ds}^2=-e^{\nu(r)}{dt}^2+e^{-\nu(r)}{dr}^2+r^2{d\theta}^2+r^2\sin^2(\theta){d\phi}^2
\end{equation}

In PGT , torsion must also satisfy the Killing equation $L_{\xi}
T_{\mu\nu}^{\rho}=0$ where $L_{\xi}$ is the Lie derivative in the
direction of $\xi$. By using this condition, the explicit form of
the torsion tensor for a static spherically symmetric spacetime can
be written as \cite{spain,Rauch2,Sur}

$${T^t}_{tr}=-{T^t}_{rt}=a(r)\,,\,{T^r}_{\theta\phi}=-{T^r}_{\phi\theta}=k(r)\sin(\theta)e^{\nu(r)}$$
$${T^r}_{tr}=-{T^r}_{rt}=a(r)e^{\nu(r)}\,,\,{T^t}_{\theta\phi}=-{T^t}_{\phi\theta}=k(r)\sin(\theta)$$
$${T^\phi}_{t\theta}=-{T^\phi}_{\theta t}=\frac{h(r)e^{\nu(r)}}{\sin(\theta)}\,,\,{T^\theta}_{r\phi}=-{T^\theta}_{\phi r}=h(r)\sin(\theta)$$
$${T^\phi}_{\theta r}=-{T^\phi}_{r\theta}=\frac{h(r)}{\sin(\theta)}\,,\,{T^\theta}_{\phi t}=-{T^\theta}_{t\phi}=h(r)\sin(\theta)e^{\nu(r)}$$
$${T^\theta}_{\theta t}=-{T^\theta}_{t\theta}={T^\phi}_{\phi t}=-{T^\phi}_{t\phi}=g(r)e^{\nu(r)}$$
\begin{equation}\label{Tcomps}
    {T^\theta}_{r\theta}=-{T^\theta}_{\theta r}={T^\phi}_{r\phi}=-{T^\phi}_{\phi r}=g(r)
\end{equation}

 where $a(r),k(r),h(r)$ and $g(r)$ are four unknown functions. Using
 (\ref{metric}), (\ref{Tcomps}) and the field equations (\ref{field
 eq1}) and (\ref{field eq2}), we get the following differential
 equations

$$\nu^\prime(r)\Big[\Big(4g(r)-2a(r)-\nu^\prime(r)-\frac 4 r\Big)\nu^\prime(r)$$
$$+\frac 1 r\Big(\frac 2
    r-4a(r)+8g(r)\Big)-4a^\prime(r)+8g^\prime(r)-3\nu^{\prime\prime}(r)\Big]$$
$$+2\Big(2g(r)-a(r)-\frac
    2 r\Big)\nu^{\prime\prime}(r)$$
$$+\frac 4
    r\Big[2g^\prime(r)-a^\prime(r)+\frac 1 r\Big(a(r)-2g(r)\Big)$$
\begin{equation}\label{eq11}
    +\frac 4
    {r^3}\Big(1-e^{-\nu(r)}\Big)-\nu^{\prime\prime\prime}(r)-2a^{\prime\prime}(r)+4g^{\prime\prime}(r)\Big]=0
\end{equation}

$$a_0\Big(r\nu^{\prime}(r)+1-e^{-\nu(r)}\Big)$$
$$+\frac 1
    {24r^4}\Big\{be^{\nu(r)}\Big[\Big(4g(r)-2a(r)-\nu^\prime(r)-\frac 4
    r\Big)r^2\nu^\prime(r)$$
$$+r^2\Big(-2a^\prime(r)+4g^\prime(r)-\nu^{\prime\prime}(r)+\frac
    8 rg(r)-\frac 4
    ra(r)\Big)$$
$$-2+2e^{-\nu(r)}\Big]\times\Big[\Big(\nu^\prime(r)+2a(r)\Big)r^4\nu^\prime(r)$$
$$+r^4\Big(4g^\prime(r)+\nu^{\prime\prime}(r)+2a^\prime(r)-8g(r)a(r)+\frac
    8 rg(r)+\frac 4
    ra(r)\Big)$$
\begin{equation}\label{eq200}
    +r^2\Big(4h(r)k(r)-2+2e^{-\nu(r)}\Big)+2{k(r)}^2\Big]\Big\}=0
\end{equation}

$$\Big[\Big(4g(r)-2a(r)-\nu^\prime(r)-\frac 4
    r\Big)\nu^\prime(r)-2a^\prime(r)+4g^\prime(r)$$
$$-\nu^{\prime\prime}(r)+\frac
    8 rg(r)-\frac 4 ra(r)-\frac 2
    {r^2}\Big(1-e^{-\nu(r)}\Big)\Big]$$
$$\times\Big[2g^\prime(r)-4g(r)a(r)+\frac 2
    r\Big(a(r)+g(r)\Big)$$
\begin{equation}\label{eq201}
    +\frac {k(r)} {r^2}\Big(\frac {k(r)} {r^2}+2h(r)\Big)\Big]=0
\end{equation}

$$a_0\Big(r\nu^\prime(r)+1-e^{-\nu(r)}\Big)$$
$$+\frac 1
{24r^4}\Big\{be^{\nu(r)}\Big[\Big(4g(r)-2a(r)-\nu^\prime(r)-\frac 4
    r\Big)r^2\nu^\prime(r)$$
$$+r^2\Big(-2a^\prime(r)+4g^\prime(r)-\nu^{\prime\prime}(r)+\frac
    8 rg(r)-\frac 4
    ra(r)\Big)$$
$$-2+2e^{-\nu(r)}\Big]\times\Big[\Big(\nu^\prime(r)+2a(r)\Big)r^4\nu^\prime(r)$$
$$+r^4\Big(-4g^\prime(r)+\nu^{\prime\prime}(r)+2a^\prime(r)+8g(r)a(r)-\frac
    4 ra(r)\Big)$$
\begin{equation}\label{eq211}
    +r^2\Big(-4h(r)k(r)-2+2e^{-\nu(r)}\Big)-2{k(r)}^2\Big]\Big\}=0
\end{equation}

$$a_0\Big[\Big(\nu^\prime(r)+\frac 2 r\Big)\nu^\prime(r)+\nu^{\prime\prime}(r)\Big]$$
$$-\frac b 3e^{\nu(r)}\Big[\Big(4g(r)-2a(r)-\nu^\prime(r)-\frac
4r\Big)\nu^\prime(r)-2a^\prime(r)$$
$$+4g^\prime(r)-\nu^{\prime\prime}(r)+\frac8 rg(r)-\frac 4 ra(r)-\frac 2 {r^2}(1-e^{-\nu(r)})\Big]$$
$$\times\Big[\frac {g(r)} r-\frac 1 {2r^2}\Big(1-e^{-\nu(r)}\Big)+\frac 1 2\nu^\prime(r)\Big(a(r)+\frac 1 2\nu^\prime(r)\Big)$$
\begin{equation}\label{eq222}
    +\frac 1 2a^\prime(r)+\frac 1 4\nu^{\prime\prime}(r)\Big]=0
\end{equation}

Where a prime denotes differentiation with respect to $r$. After
solving the differential equations analytically, we can obtain the
solution for 5 unknown functions $\nu(r)$, $a(r)$, $g(r)$, $h(r)$
and $k(r)$. However equations (\ref{eq200}) and (\ref{eq211}) are
not independent of each other. This leads to a dependent solution
for torsion functions $h(r)$ and $k(r)$. Torsion solutions of the
system are presented in the appendix. The solution for the metric
function $\nu(r)$ is

\begin{equation}\label{sol1}
     \nu(r)=\ln\Big(1-\frac {c_1} r+\frac {c_2} {r^2}\Big)
\end{equation}

Substituting this relation in (\ref{metric}), we get the equivalent
to the Schwarzschild metric in PGT. For positive values of constant
$c_2$, this is similar to the Reissner-Nordstrom solution in general
relativity. This result is consistent with the results of reference
\cite{spain}. It should be noted that our solutions are obtained for
general values of coupling constants $a_0$, $a_{1}$ and $b$. The
solutions have 3 constants of integration $c_1$, $c_2$ and $c_3$
(The constant $c_3$ appears in the solutions for torsion functions
given in the appendix). For interpreting these constants, we
consider a limiting case of the differential equations (17-21) when
all components of the torsion tensor are set to zero. If we consider
this particular case, $a(r)=h(r)=g(r)=k(r)=0$, then equation
(\ref{eq201}) becomes trivial and we have a set of differential
equations just in $\nu(r)$. By solving equation (\ref{eq11}) for
$\nu(r)$ in this case, we obtain the following form

\begin{equation}\label{solGR}
     \nu(r)=\ln\Big(1-\frac {c_4} r+\frac {c_5} {r^2}-\frac {c_6r^2} {12}\Big)
\end{equation}

But this solution must satisfy all other equations. Substituting
(\ref{solGR}) in equations (18), (20) and (21) , and solving for the
contstants  $c_4$, $c_5$ and $c_6$, gives $c_5=c_6=0$. In this case,
as one can see, the Schwarzschild metric of GR will be recovered
from (\ref{solGR}).  Combining these results, we find that the
constant $c_1$ is related to the mass of the black hole. Also
constants $c_2$ and $c_3$ should be related to the spin effects
which induce torsion in the spacetime. From now on we write the
solution (\ref{sol1}) for $\nu(r)$ as

\begin{equation}\label{sol}
    \nu(r)=\ln\Big(1-\frac {2m} r+\frac S {r^2}\Big)
\end{equation}

In which $m$ and $S$ are some charges related to field strengths of
curvature and torsion, respectively. The black hole metric
(\ref{metric}) then will be

\begin{equation}\label{metricbh}
    {ds}^2=-f(r){dt}^2+\frac 1 {f(r)}{dr}^2+r^2{d\theta}^2+r^2\sin^2(\theta){d\phi}^2
\end{equation}

\begin{equation}\label{fr}
f(r)=1-\frac {2m} r+\frac S {r^2}
\end{equation}

The roots of $f(r)$ are  black hole horizons
\begin{equation}\label{horr}
R_\pm=m\pm\sqrt{m^2-S}
\end{equation}

with the following condition

 \begin{equation}\label{conditionbh}
    m^2\geq S
\end{equation}

$R_+$ is outer horizon and can be interpreted as the Schwarzschild
radius of the black hole. From now on we rename   $R_+$ as $R$. For
the positive values of $S$, we have $R_{PGT}<R_{GR}$ while for
negative values of $S$, the opposite is true \emph{i.e.}
$R_{PGT}>R_{GR}$. As expected, in the limiting case of $S=0$, the
Schwarzschild radius is equal in  Poincar{\'e} gauge theory and
general relativity .

\section{Effective Potential and orbits for a Black hole with torsion }

For the line element (\ref{metricbh}), there exist two killing
vectors $\xi$ associated with energy $E$ and the angular momentum
$L$ per unit mass which are conserved quantities of motion along the
geodesics
\begin{equation}\label{kilE}
    \xi_\mu=\Big(-(1-\frac {2m} r+\frac S {r^2}),0,0,0\Big)
\end{equation}

\begin{equation}\label{kilL}
    \xi_\mu=\Big(0,0,0,r^2\sin^2(\theta)\Big)
\end{equation}

 In equatorial plane $\theta=\frac \pi 2$ the line element
is simplified as follows

\begin{equation}\label{metric2}
    {ds}^2=-f(r){dt}^2+\frac 1 {f(r)}{dr}^2+r^2{d\phi}^2
\end{equation}

We also have for E and L

\begin{equation}\label{E}
    E=-P_t=-g_{tt}P^t=f(r)\frac {dt} {d\lambda}\rightarrow \frac {dt}
    {d\lambda}=\frac E {f(r)}
\end{equation}

\begin{equation}\label{L}
    L=P_\phi=g_{\phi\phi}P^\phi=r^2\frac {d\phi} {d\lambda}\rightarrow \frac {d\phi}
    {d\lambda}=\frac L {r^2}
\end{equation}

Where P is energy-momentum $4$-vector and $\lambda$ is an affine
parameter.

With substituting relations (\ref{E}) and (\ref{L}) in line element
(\ref{metric2}) and some straightforward calculations we obtain
relation as follows

\begin{equation}\label{metric2LE}
    k={(\frac {ds} {d\lambda})}^2=-\frac {E^2} {f(r)}+\frac 1 {f(r)}{(\frac {dr}
    {d\lambda})}^2+\frac {L^2} {r^2}
\end{equation}

where $k$ is equal to $-1$ , $+1$ and $0$ for timelike, spacelike
and null geodesics, respectively. Equation (\ref{metric2LE}) can be
written as

\begin{figure*}[thbp]
\begin{tabular}{rl}
\includegraphics[width=9cm]{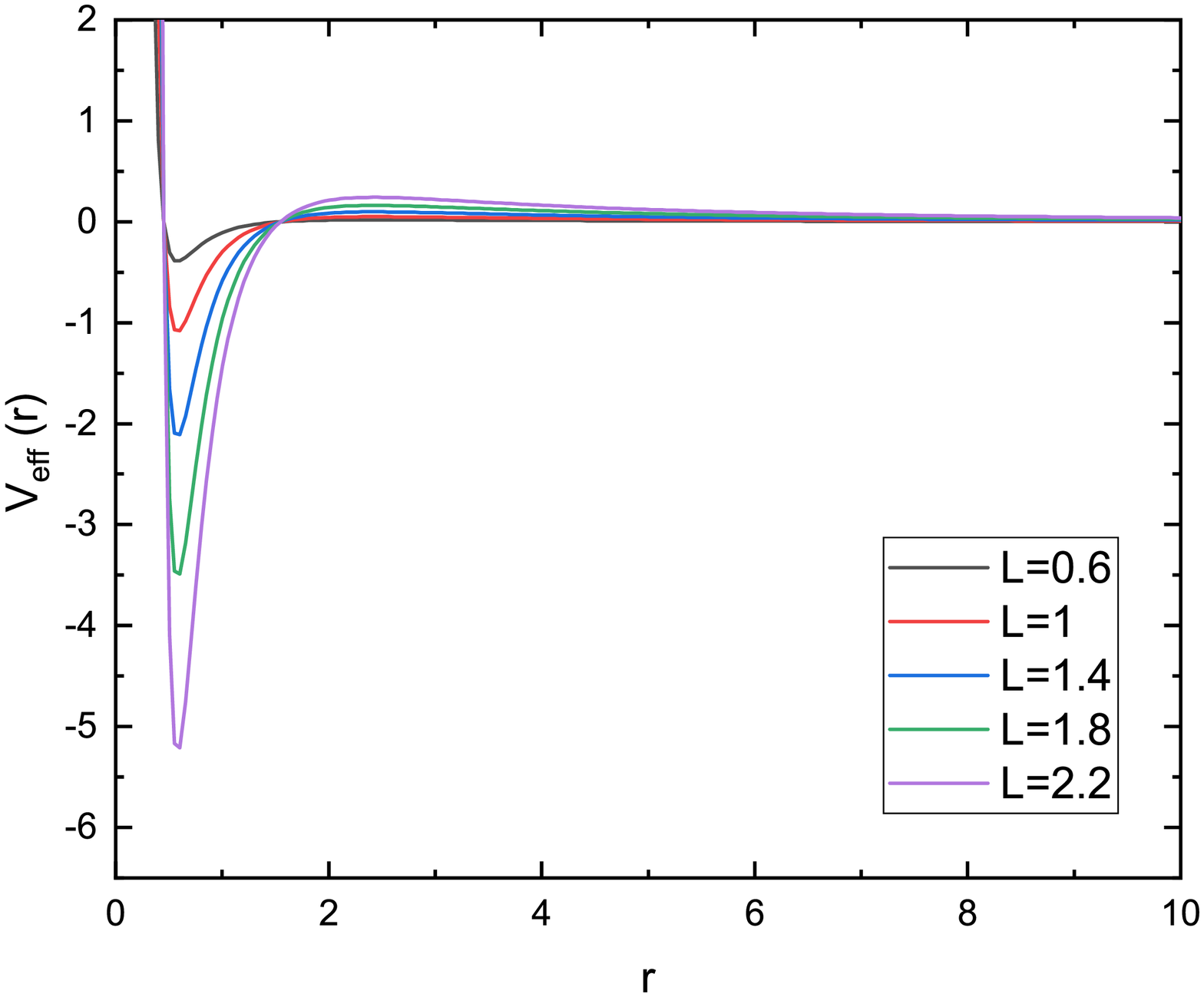}&
\includegraphics[,width=9cm]{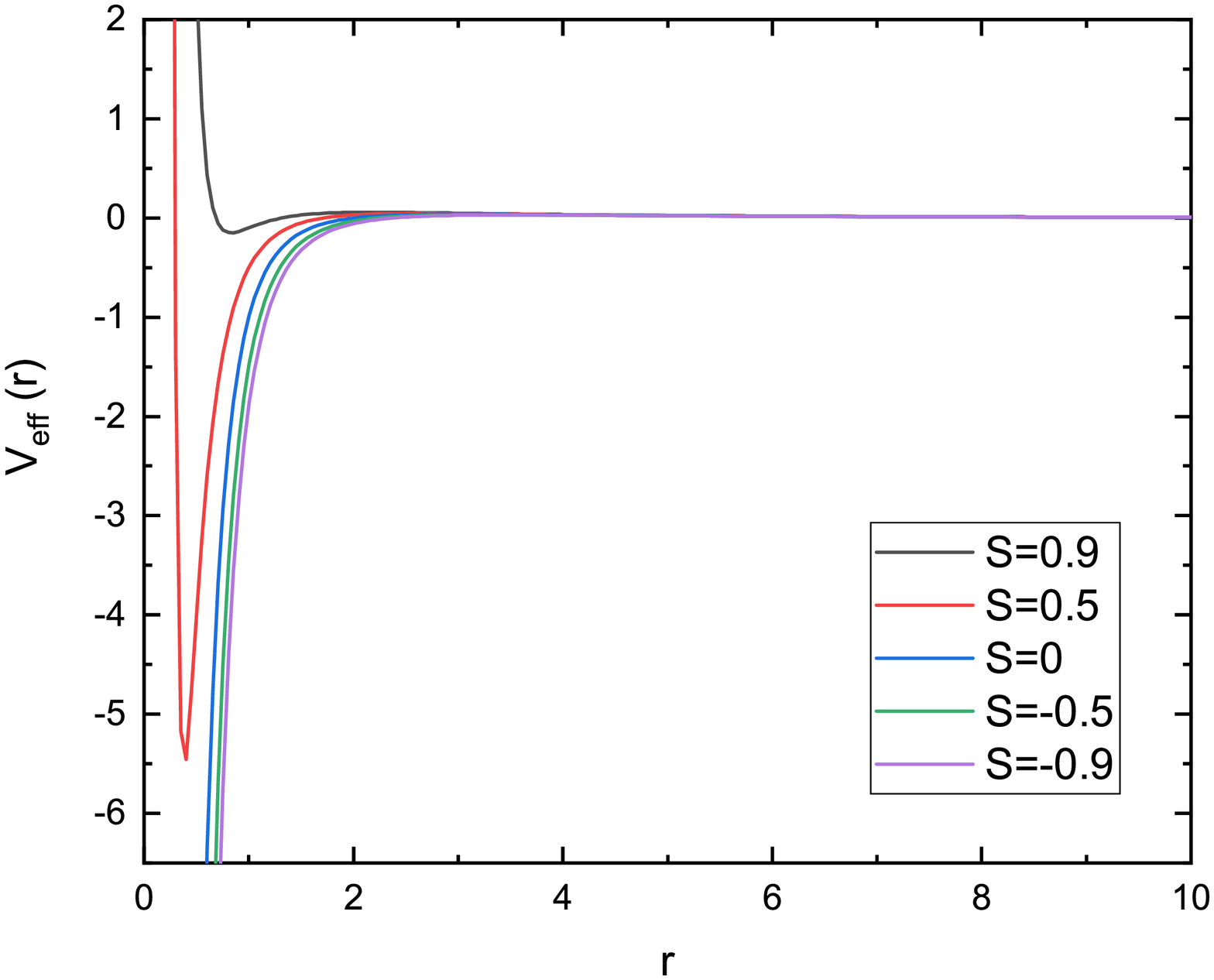}\\
\end{tabular}
\caption{\small{Behavior of effective potential of equation
(\ref{veff}) for photons ($k=0$): The left figure shows the
effective potential for different values of orbital angular momentum
per unit mass $L$ with constant $S=0.7$ and $m=1$. The right figure
shows the same effective potential for different values of $S$ with
constant $L=1$ and $m=1$.}}.
\end{figure*}

\begin{equation}\label{k}
    {\Big(\frac {dr}
    {d\lambda}\Big)}^2=E^2-\Big(1-\frac {2m} r+\frac S {r^2}\Big)\Big(-k+\frac {L^2} {r^2}\Big)
\end{equation}

From the above relation, we define the effective potential as
follows

\begin{equation}\label{veff}
 V_{eff}(r)=\Big(1-\frac {2m} r+\frac S {r^2}\Big)\Big(-k+\frac {L^2} {r^2}\Big)
\end{equation}

The position of the orbits for the effective potential is given by
the condition

\begin{equation}
\frac {dV_{eff}} {dr}\Big|_{r_c}=0
\end{equation}

The orbits are stable if
\begin{equation}
\frac {d^2V_{eff}} {dr^2}\Big|_{r_c}>0
\end{equation}

The quantities $m$ and $L^{2}$ in (\ref{veff}) are positive, however
$S$ could be positive, negative or zero. We will examine orbits of
particles for different values of $S$ and compare the result to the
orbits in general relativity ($S=0$). For photons ($k=0$), we have

\begin{equation}
\frac {dV_{eff}}
{dr}=-\frac{2L^2}{r^3}+\frac{6mL^2}{r^4}-\frac{4SL^2}{r^5}
\end{equation}

Using equation (39), we get the following solution for the position
of circular orbits around a black hole

\begin{equation}\label{r}
\frac {dV_{eff}} {dr}\Big|_{r_c}=0\,\longrightarrow\,r_{c\pm}=\frac
3 2m\pm\frac 1
    2\sqrt {9m^2-8S}
\end{equation}

This result is independent of $L$. We now look for the stability of
these two orbits. For the $r_{c+}$ we have

\begin{equation}
\frac {d^2V_{eff}} {dr^2}\Big|_{r_{c_+}}=  -64\,{\frac {{L}^{2}
\Big( 3\,m\sqrt {9\,{m}^{2}-8\,S}+9\,{m}^{2}-8\, S \Big) }{ \Big(
3\,m+\sqrt {9\,{m}^{2}-8\,S} \Big) ^{6}}}
\end{equation}

The condition (\ref{conditionbh}) ensures that $9\,{m}^{2}-8\,S$ in
the above equation will always be positive or zero; as $m$ and
$L{^2}$ are also positive or zero, $\frac {d^2V_{eff}}
{dr^2}\Big|_{r_{c_+}}$ will always be negative or zero in this case
and as a result we conclude that the orbit given by $r_{c_+}$ is not
stable. For the  $r_{c_-}$ orbit we have
\begin{equation}
\frac {d^2V_{eff}} {dr^2}\Big|_{r_{c_-}}= 64\,{\frac {{L}^{2} \Big(
3\,m\sqrt {9\,{m}^{2}-8\,S}-9\,{m}^{2}+8\,S
 \Big) }{ \Big( -3\,m+\sqrt {9\,{m}^{2}-8\,S} \Big) ^{6}}}
\end{equation}

This orbit will be stable for $ S>0 $ and unstable for $ S<0 $. In
general, there are no stable orbits of photons for negative $S$. For
the limiting case of $ S=0 $, we get

\begin{equation}
\lim _{S\rightarrow 0}r_{c_-}=0\,,\,\lim _{S\rightarrow 0}r_{c_+}=3m
\end{equation}

which is consistent with the GR case where there exist a single
unstable orbit at $r=3m$ for photons. Figure (1) shows the effective
potential of equation (\ref{veff}). One can see that the depth of
the effective potential well increases with increasing orbital
angular momentum $L$ but decreases with increasing $S$. For positive
$S$ there exist one stable orbit at $r=r_{c_-}$ (minimum of the
potential) and one unstable one at $r=r_{c_+}$. There are no stable
orbits for negative $S$ as can be seen from the right figure.

For massive particles. $k=-1$, we have

\begin{equation}
\frac {dV_{eff}}
{dr}=\frac{2m}{r^2}-\frac{2(S+L^2)}{r^3}+\frac{6mL^2}{r^4}-\frac{4SL^2}{r^5}
\end{equation}

In this case, equation (37) can be written as

\begin{equation}
 m{r_c}^{3}- \left( {L}^{2}+S \right) {r_c}^{2}+3\,m{L}^{2}r_c-2\,S{L}^{2}=0
 \end{equation}

In the limiting case of $S=0$, we have

\begin{equation}
r_c\Big(mr_c^2-L^2r_c+3mL^2\Big)=0
\end{equation}

which gives the solutions as $ r_c=0 $ and

\begin{equation}
r_{c\pm}=\frac{L^2}{2m}\Bigg(1\pm\sqrt{1-\frac{12m^2}{L^2}}\Bigg)
\end{equation}

Using the stability condition (38), we find that the $r_{c+}$ orbit
is always stable while $r_{c-}$ orbit is always unstable. In the
case of $ L^2=12m^2 $ these two orbits will coincide at $r_{c}=6m$.

For general values of $S$, equation (45) has three solutions given
by

\begin{equation}
r_{c_1}=\frac 1 {3m}\,\Big(L^2+S+\frac \alpha 2+\frac {2\beta}
{\alpha}\Big)
\end{equation}

\begin{equation}
r_{c_{2,3}}=\frac 1 {3m}\,\Big(L^2+S-\frac \alpha 4-\frac {\beta}
{\alpha}\Big)\pm i\frac{\sqrt{3}}{6m}\Big(\frac \alpha 2-\frac
{2\beta} {\alpha}\Big)
\end{equation}

where

$$\alpha=\Big[4L^2\Big(2L^4-27m^2(L^2-S)+6S(L^2+S)\Big)+8S^3$$
$$+12\sqrt3mL\Big(L^4(108m^4-126m^2S+24S^2)$$
$$+L^6(8S-9m^2)+L^2S^2(24S-9m^2)+8S^4\Big)^{\frac 1 2}\Big]^{\frac 1 3}$$

$$\beta=L^2(L^2-9m^2+2S)+S^2$$

\begin{figure*}[thbp]
\begin{tabular}{rl}
\includegraphics[width=9cm]{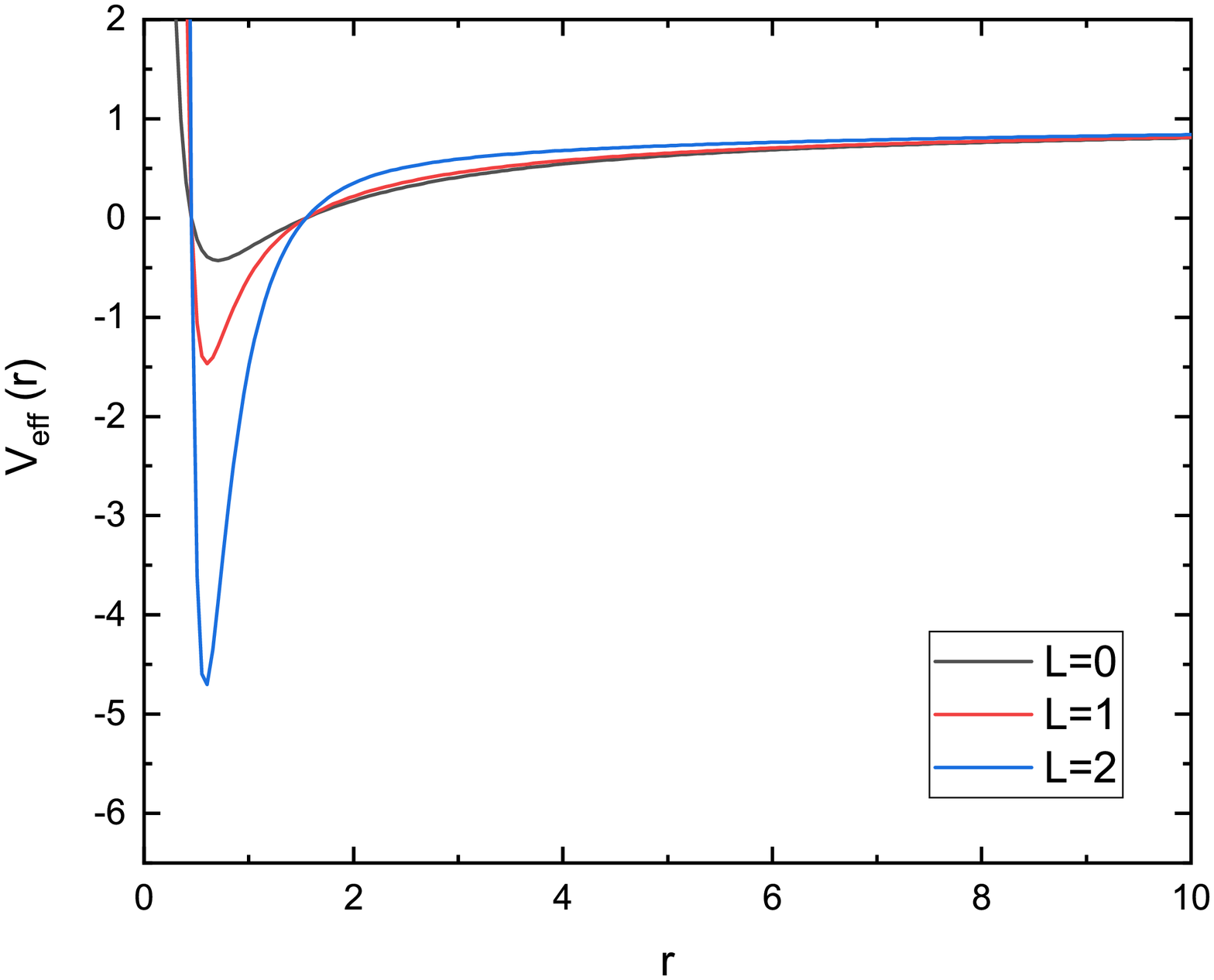}&
\includegraphics[,width=9cm]{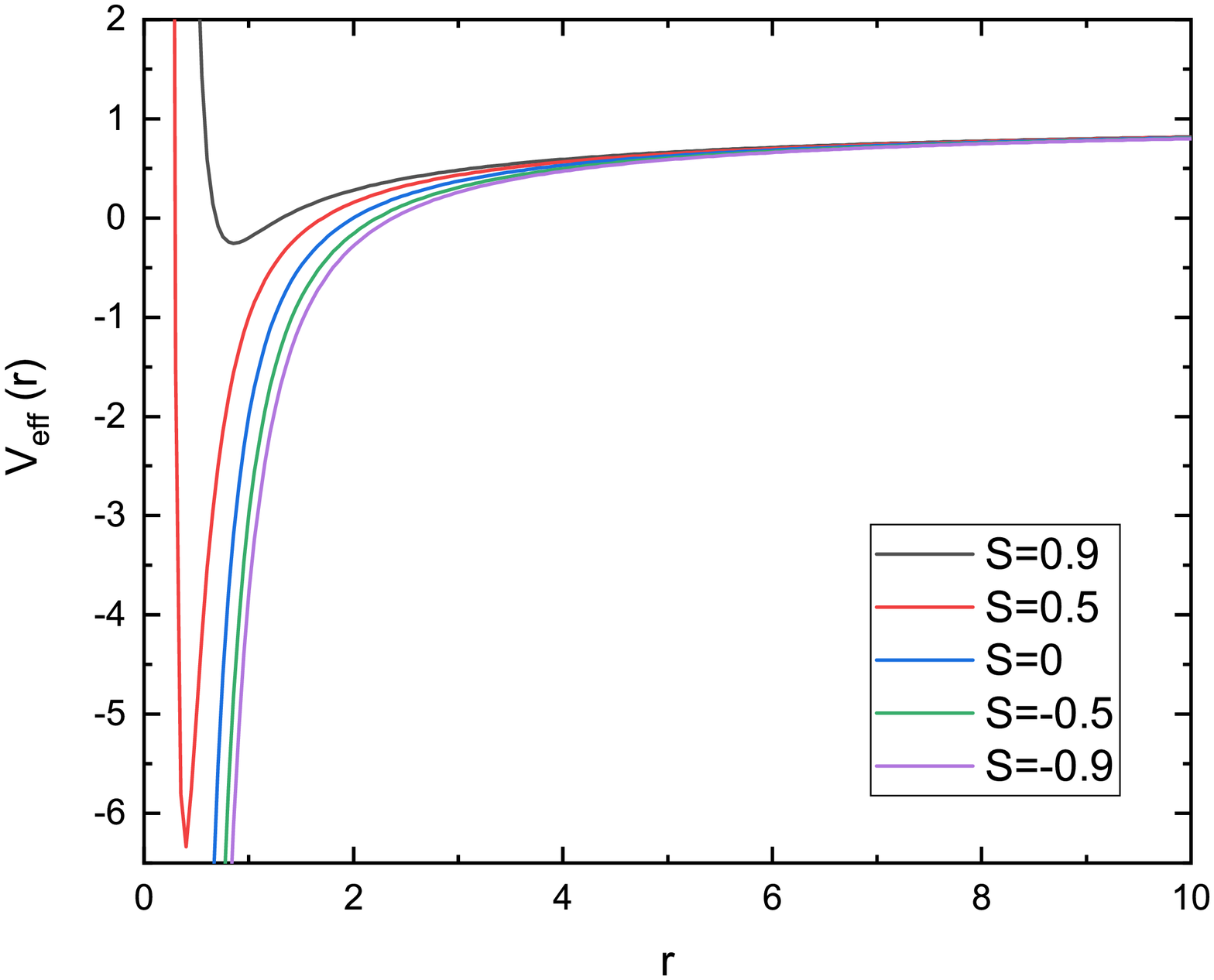}\\
\end{tabular}
\caption{\small{Behavior of effective potential of equation
(\ref{veff}) for massive particles ($k=-1$), case 1, $L^2<12m^2$:
The left figure shows the effective potential for different values
of orbital angular momentum per unit mass $L$ with constant $S=0.7$
and $m=1$. The right figure shows the same effective potential for
different values of $S$ with constant $L=1$ and $m=1$.}}
\end{figure*}

\begin{figure*}[thbp]
\begin{tabular}{rl}
\includegraphics[width=9cm]{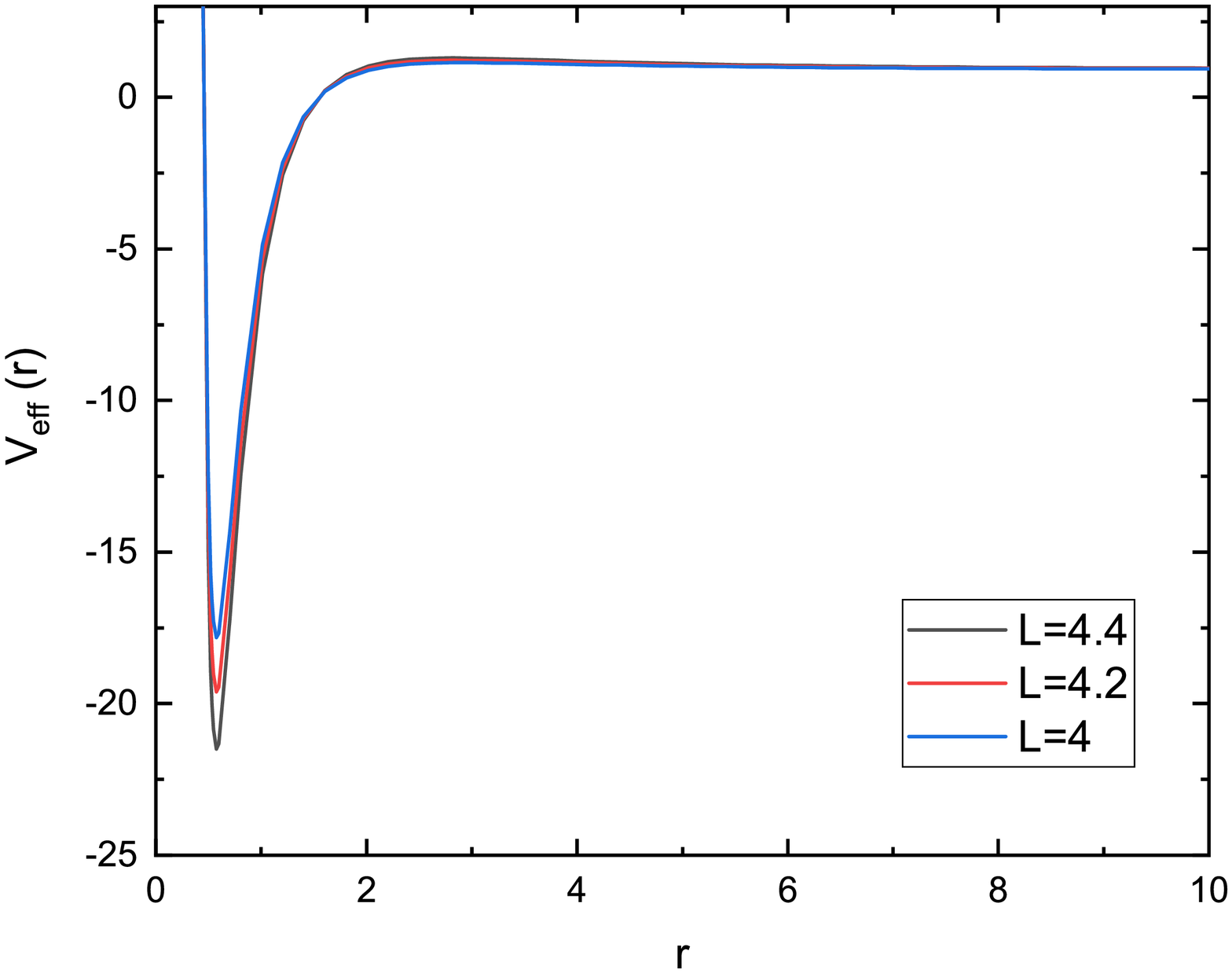}&
\includegraphics[,width=9cm]{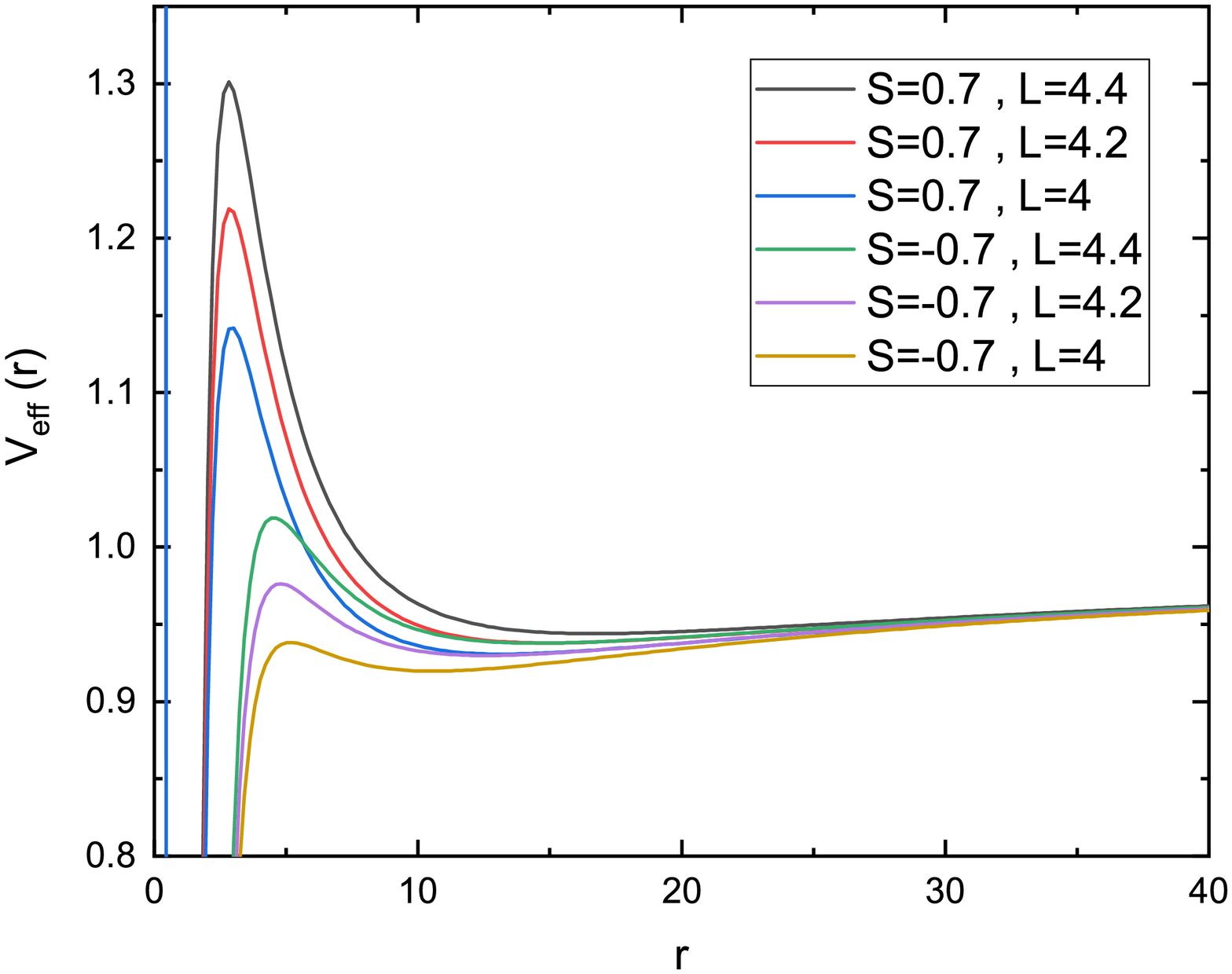}\\
\end{tabular}
\caption{\small{ Behavior of effective potential of equation
(\ref{veff}) for massive particles ($k=-1$), case 2, $L^2>12m^2$:
The left figure shows the effective potential for different values
of orbital angular momentum per unit mass $L$ with constant $S=0.7$
and $m=1$. The right figure shows different part of the same
potential for $S=0.7$ and $S=-0.7$.}}
\end{figure*}

\begin{figure*}[thbp]
\begin{tabular}{rl}
\includegraphics[width=9cm]{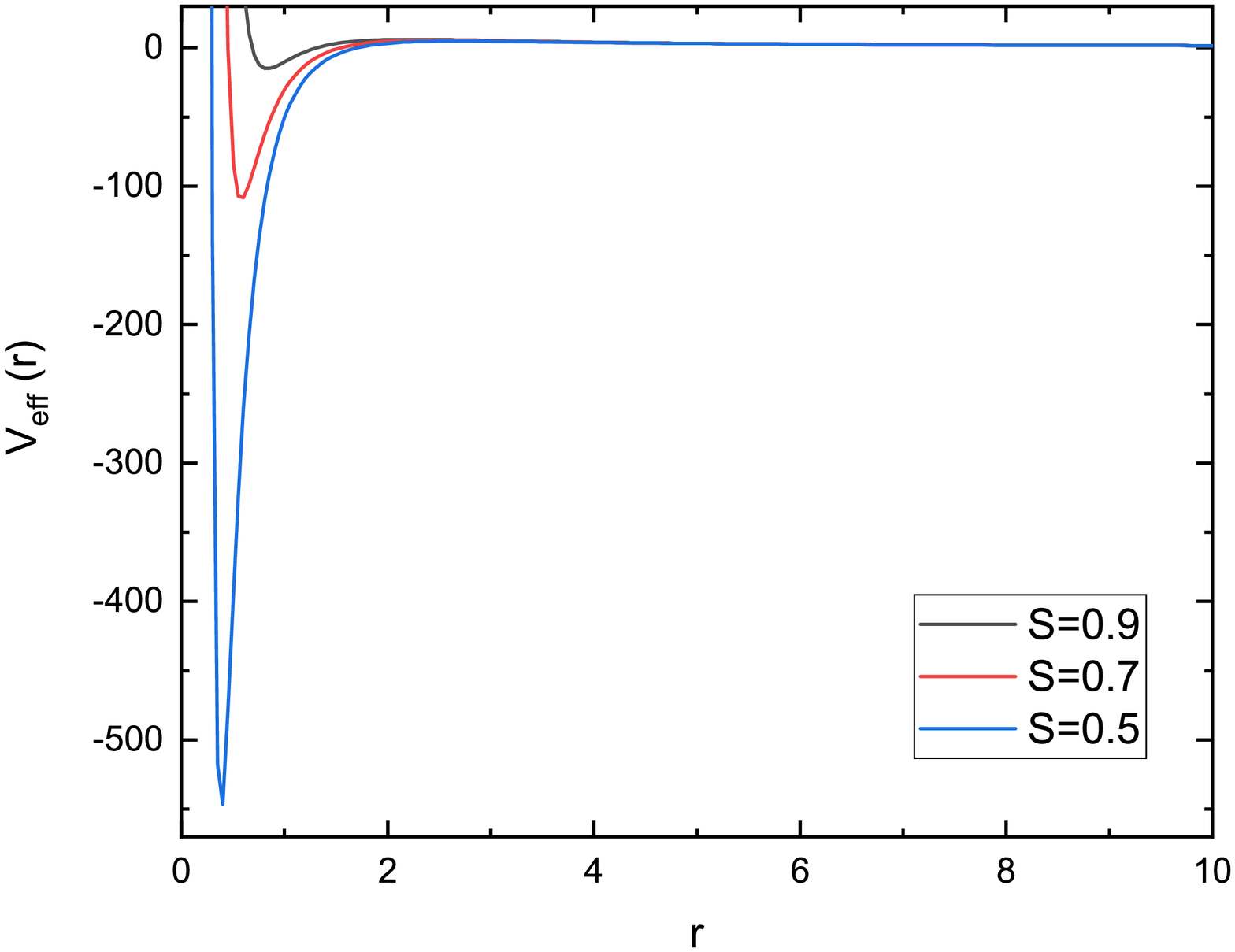}&
\includegraphics[,width=9cm]{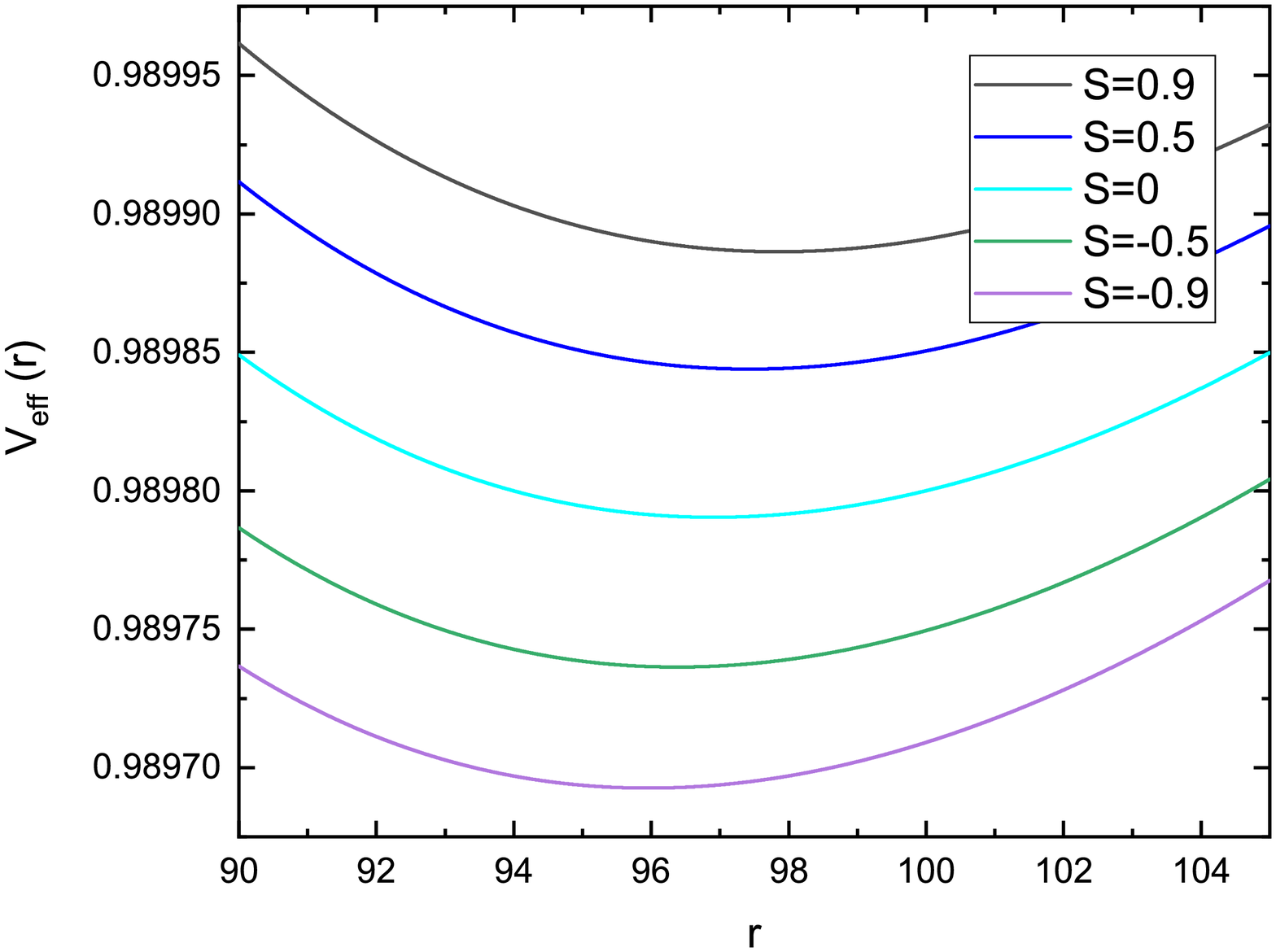}\\
\end{tabular}
\caption{\small{Behavior of effective potential of equation
(\ref{veff}) for massive particles ($k=-1$), case 2, $L^2>12m^2$:
The left figure shows the effective potential for different values
of $S$ with constant $L=10$ and $m=1$. The right figure shows
different part of the same potential}}
\end{figure*}

In order to determine the sign of the above solutions, we employ the
Cardano method for solving cubic equations. Using the following
change of parameter

\begin{equation}
r_c=x+{\frac {{L}^{2}+S}{3m}}
\end{equation}

equation (45) can be rewritten as

\begin{equation}
 x^3+px+q=0
\end{equation}

where

 \begin{equation}
 p=\frac{L^2(9m^2-L^2-2S)-S^2}{3m^2}
 \end{equation}

 \begin{equation}
q=\frac{L^2(27m^2L^2-2L^4-6L^2S-27m^2S-6S^2)+2S^3}{27m^3}
 \end{equation}

 To determine whether the solutions are real or complex and also the
 sign of the solutions, we define

\begin{equation}
\Delta={\Big(\frac{p}{3}\Big)}^3+{\Big(\frac{q}{2}\Big)}^2
\end{equation}

Then, if $\Delta>0$, there exist a single real solution to equation
(51). In the case of $\Delta<0$ there are three real solutions.
Finally if $\Delta=0$, there are three real solutions, two of which
coincide.

Substituting (52) and (53) in  (54), we get for the case $S=0$

\begin{equation}
\Delta=-\frac{L^6}{12m^2}\,\Big[L^2-12m^2\Big]
\end{equation}

The sign of $\Delta$ in (55) depends on the sign of the term inside
the brackets. Interestingly this term is what determines the
existence and stability of solutions in genera relativity, as
obvious from equation (47). Motivating by this, we analyze the
solutions of equation (45) for general $S$ in three different
cases:\\

$\bullet$ Case 1: For $L^2<12m^2$ there exist a single stable orbit
for $S>0$ (Figure (2)). The radius of this orbit increases with increasing $S$. There can be no orbits for $S<0$, as one can see in the right figure.\\

$\bullet$ Case 2: For $L^2>12m^2$,  there exist two stable orbits
and one unstable orbit for $S>0$ (Figure (3)). For $S<0$ there exist
one stable and one unstable orbit. The radius of stable orbits
increase with increasing $S$  for all values of $S$. The opposite is
true for unstable orbits. These stable orbits and their behaviors
can also be seen in figure (4) for different values of $S$.
This figure shows that the radius of second stable orbit increases with increasing $L$. \\

$\bullet$ Case 3: For $L^2=12m^2$, there exist two stable orbits and
one unstable orbit for $S>0$. There are no orbits for $S<0$. The
behavior of orbits is the same as the case 2.\\

It is also interesting to examine the case of particles with zero
orbital angular momentum per unit mass, $L=0$. In this case we have

\begin{equation}
\lim _{L\rightarrow 0}r_{c_1}=\frac{S}{m}
\end{equation}

and the other solutions vanish in this limit. The stability
condition takes the form

\begin{equation}
\lim _{L\rightarrow 0}{\Big(\frac{d^2V_{eff}}
{dr^2}|_{r_{c1}}\Big)}=\frac{2m^4}{S^3}
\end{equation}

which is always greater than zero for $S>0$. This is an interesting
result, as in GR there are no orbits for $L=0$. As a consequence,
the stable orbit for $S>0$ given by (49) for spacetime with torsion,
is a result of the interaction between the spin of particle and the
background spacetime.

\section{Black hole thermodynamics with torsion}

We now turn our attention to black hole thermodynamics in the
presence of torsion. We begin with the definition of the surface
gravity

\begin{equation}\label{sg}
\kappa=\frac 1 2f^\prime(R)
\end{equation}

where $f(r)$ and the horizon radius $R$ now are given by equations
(\ref{fr}) and (\ref{horr}) respectively. Using (\ref{fr}) and
(\ref{sg}), we get

\begin{equation}
\kappa=\frac {\sqrt{m^2-S}} {({m+\sqrt{m^2-S}})^2}
\end{equation}

In the GR limit, $ S=0 $, this equation takes the form

\begin{equation}
\lim _{S\rightarrow 0}{\frac {\sqrt{m^2-S}}
{({m+\sqrt{m^2-S}})^2}}=\frac{1}{4m}=\kappa_{GR}
\end{equation}

One can see that in the appropriate limit, the surface gravity
approaches its GR value. In order to obtain the first law of black
hole thermodynamics in the presence of torsion, we consider a black
hole with parameters  $ m $ and $ S $ and assume that the black hole
undergoes a change in the parameters by a quasi-static process to
new parameter values $m+\delta m$ and $S+\delta S$. The surface area
of the horizon, $A=4\pi R^2$ is a function of the parameters  $ m $
and $ S $ by the virtue of equation (\ref{horr}).

\begin{equation}
A=4\,\pi \, \left( m+\sqrt {{m}^{2}-S} \right) ^{2}
\end{equation}

A change in this parameters gives

\begin{equation}
\delta A=\frac {\partial A} {\partial m}\,\delta m+\frac {\partial
A} {\partial S}\,\delta S
\end{equation}

Combining the last two equations we get

\begin{equation}
\delta A=\frac{8\,\pi \, \left( m+\sqrt {{m}^{2}-S} \right)^2}{\sqrt
{{m}^{2}-S}} \,\delta m-\,{\frac {4\pi \, \left( m+\sqrt {{m}^{2}-S}
\right) }{\sqrt {{m}^{2} -S}}}\,\delta S
\end{equation}

A simple algebra gives

\begin{equation}\label{deltam}
    \delta m=\frac {\sqrt{m^2-S}} {2\pi({m+\sqrt{m^2-S}})^2}\,\frac {\delta A}
    {4}+\frac {\delta S} {2(m+\sqrt{m^2-S})}
\end{equation}

We define the temperature $T$ and entropy $\Sigma$ of the black hole
in the presence of torsion by

\begin{equation}
T=\frac {\sqrt{m^2-S}} {2\pi({m+\sqrt{m^2-S}})^2}\,\,,\,\,\Sigma=\pi
\, \left( m+\sqrt {{m}^{2}-S} \right) ^{2}
\end{equation}

If we compare equation (\ref{deltam}) with the general law of black
hole thermodynamics

\begin{equation}\label{first law}
    \delta m=\frac 1 {8\pi} \kappa\delta A+\Omega\delta J+\Phi\delta Q
\end{equation}

we can see that the first term in the right-hand-side of equation
(\ref{deltam}) corresponds to the $Td\Sigma$ term in the first law
of classical thermodynamics and also to the $\frac 1 {8\pi}
\kappa\delta A$ term in equation (\ref{first law}). For a static
black hole, the second term in (\ref{first law}) vanishes. Also if
we define torsion 'charge' and 'potential' as

\begin{equation}
Q_{Torsion}=S \,\,,\,\,{\Phi}_{Torsion}={\frac {1}
{2(m+\sqrt{m^2-S})}}
\end{equation}

 we can interpret the second term in the right-hand-side of equation
(\ref{deltam}) as the $\Phi\delta Q$ term in (\ref{first law}). This
definition of the torsion potential is also consistent with
(\ref{horr}).

\section{CONCLUSION}

In this paper, we study static spherically symmetric solutions to
the field equations in the Poincar{\'e} gauge theory of gravity. The
effects of torsion appear as a single parameter in the line element.
For the positive values of this parameter, the resulting line
element is found to be of the Reissner-Nordstrom type and has two
distinct horizons. This result is in agreement with the results of
\cite{spain}. The effects of torsion appear as a sort of torsion
'charge' related to the spin of matter.   We also study particle
orbits around a black hole in this geometry for both massless and
massive particles. For massless particles, there exist one stable
orbit at $r=r_{c_-}$ and one unstable one at $r=r_{c_+}$, where
$r=r_{c_\pm}$ are given by equation (\ref{r}). There are no stable
orbits for negative $S$. For massive particles, there exist one
stable orbit for $S>0$ and no orbit for $S<0$ in the case
$L^2<12m^2$. For the case $L^2>12m^2$, there exist two stable orbits
and one unstable orbit for $S>0$. Also, there is one stable and one
unstable orbit for $S<0$. The case $L^2=12m^2$ is the same as $S>0$
for the case $L^2>12m^2$. There is no orbit for $S<0$. Remarkably
for massive particles, there also exist a stable orbit even when
orbital angular momentum per unit mass $L$ is set to zero. This
suggests that this orbit is a result of the interaction between spin
of particles with the torsion of the background geometry. Finally,
we derive the first law of black hole thermodynamics by defining the
temperature and entropy of a black hole with torsion in terms of its
mass and torsion charge.

\section{appendix}
In this appendix we present the full solutions to the torsion
function in the system of equations given by (17-21).

$$a(r)=\frac {{(1-\frac {c_1} r+\frac {c_2} {r^2})}^{-1}} {2r^2}\,\Big(c_1(r^3-1)\,-\frac {2a_0r}
{c_1b}\,(1-\frac {c_2} {r^2})$$

\begin{equation}\label{sola}
   +2r(1+2c_3)-2c_2\,(1-\frac 1 r)\Big)
\end{equation}

$$g(r)=\frac {{(1-\frac {c_1} r+\frac {c_2} {r^2})}^{-1}} {4r^2}\,\Big(-c_1(r^3+2)-\frac {2a_0r} {c_1b}\,(1-\frac {c_2}
    {r^2})$$
\begin{equation}\label{solg}
    +2r(1+2c_3)-2c_2(2-\frac 1 r)\Big)
\end{equation}

$$h(r)\Big[r^3+c_1r(c_1-2r)+c_2\Big(2r(c_2-c_1)+c_2\Big)\Big]=$$
$$ \Big\{2{k(r)}^2{(c_1br)}^2\Big[r^2+c_1(c_1-2r)+{(\frac {c_2}
    r)}^2+2c_2(1-\frac {c_1}
    r)\Big]$$
$$+2r{(c_1br)}^2\Big[c_1c_2r(r^3-r^2-2)+{c_1}^2r^4(1+\frac {r^3}
    2)$$
$$+2c_2(c_2(1-2r)+r^2(1+2c_3))-4c_3r^3(1+2c_3)\Big]$$
$$ +4c_1a_0br^3\Big[c_2\Big(c_1+c_2(3-\frac
    1 r-2r(1+2c_3)-3r^2)+r^3\Big)\Big]$$
\begin{equation}\label{solhk}
 +4{a_0}^2r^2\Big(c_2(2r^2-c_2)-r^4\Big)\Big\}\times\Big(-\frac 1
    {4rk(r){(c_1br)}^2}\Big)
\end{equation}\\

\end{document}